# Design Based Teaching
# for Science and Engineering Students

**Yousef R. Shayan**
Concordia University, Montréal, Québec
yshayan@ece.concordia.ca

**Mouhamed Abdulla**
Concordia University, Montréal, Québec
m_abdull@ece.concordia.ca

**Abstract**

*In this paper, we will explain a successfully proven lecturing method based entirely on system design. Wireless communications will be used as an example to explain this new teaching philosophy.*

## 1 Introduction

A society's classification as an advanced nation seems to always be dependent on its economic growth and technological advancement. And an important force for such advancement is very much dependent on engineering design, innovation and creativity.

In fact, Canadian universities, industries, and governmental bodies have helped immensely to ensure a continued prosperity in design innovation. However, more effort is still required, mainly by academics, to prepare the future generation of engineers for tomorrow's challenges.

In this paper, we hope to discuss and share our vision of how engineering system design could be taught in a classroom environment as an alternative to classical theory based teaching. This method has now been used for several years in our university for both the digital and the wireless communication courses. And the majority of students, undergraduates and graduates alike, have expressed positive and encouraging feedback which eventually led to professor Shayan receiving the Teaching Excellence Award for the 2007-2008 academic year.

Also, we must note that throughout the paper, we will use wireless communications as an example to explain our teaching concept. However, this should not stop instructors in any science or engineering field to benefit from this idea, as long as the philosophical approach and structure remains more or less uniform.

## 2 The Big Picture

It is absolutely extraordinary to ponder and reflect on how we as humans have progressed so much in so little time. In today's world there are so many scientifically related fields, subfields, and specializations that at times it is hard for us to even know where we fit in the larger scheme of things. And engineering students are no exception.

From experience, we have noticed that students often appreciate and perform quite well academically once they have a clear vision and a defined goal. A simple way to guide students toward a well defined objective would be for example to use the simplified, yet unifying science tree of Figure 1. Notice that in our case of interest, the path to telecommunications and signal processing is highlighted.

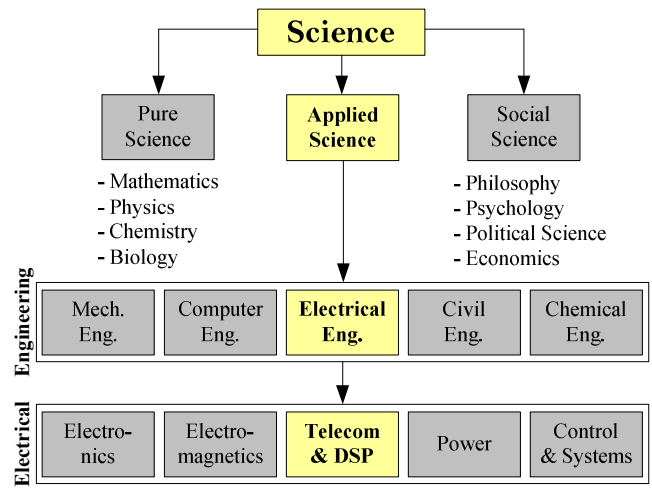

Figure 1. A unifying science tree

For junior undergraduate students, the separation of the various engineering fields into specific disciplines is essential in order to plan their future early and accordingly. On the other hand, for senior undergraduate and graduate students, showing the interdependencies of the fields in a design and manufacturing context might be more insightful for what awaits them as professionals. An example of the later, for the design of a mobile phone, is explained in Figure 2.

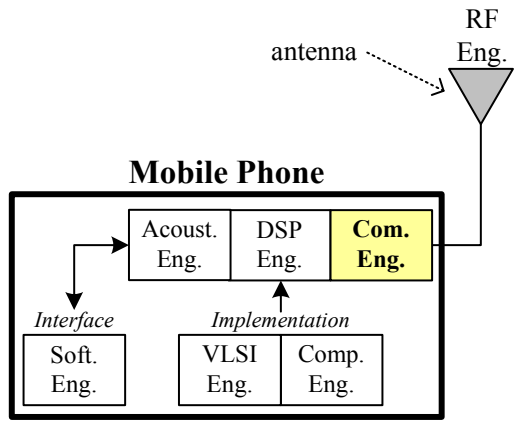

Figure 2. Example of engineering interdependencies

At first, it's quite natural to think that only a communication engineer might suffice to work on its design, because after all a cell-phone is a telecom device. But in truth, large companies have multiple specialists with variant backgrounds aiming to design only part of the mobile. And obviously, this is done while insuring a global objective of harmony and success among the other aspects of the design process.

To briefly breakdown Figure 2, the acoustic engineer deals with the microphone part of the device. The goal here is to change a sound wave into an electrical signal. Once this is done, the Digital Signal Processing [DSP] engineer will work on changing the analog signal to a digital one, and additionally insure proper frequency filtering is applied to eliminate the low power section. Then, the digital signal is passed to the communication engineer, where specific protocols are assigned, redundancy bits are added for protection, and the signal is modulated for transmission. At this point, the Radio Frequency [RF] engineer takes the electrical signal and changes it to an Electromagnetic [EM] wave using the antenna. As for the implementation, it is based on solid-state silicon chip prepared and designed by a Very Large Scale Integration [VLSI] process engineer. And the way transistors and logic units are configured and connected within the chip is usually the work of a computer engineer using low-level hardware programming language. Last, and clearly not least, a software engineer with the help of high-level coding design's the interface so that end-users enjoy easily the functionality of their device.

Here, we just witnessed an example of how up to seven specialists had to work closely to make the mobile possible because of the strong interdependencies that exit between the fields.

## 3 Traditional Design Approach

A common practice in engineering is to define a system, and then model it as best as possible. Usually we could think of a system as a machine that takes input information for processing, and then releases an output. Figure 3, shows a block diagram of a multiple input multiple output system.

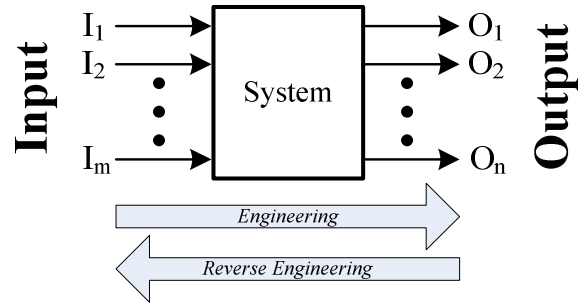

Figure 3. Traditional system design

So in traditional academic system teaching the instructor formulates a design question. And within the statement some information are given. It could either be the input or the output or even the system itself. And based on what is given, we're either doing engineering or reverse engineering manoeuvres. Sometimes it is not as clean cut as explained here, and only part of the input is provided with a subset of either the output or the system. In other instances, conflicting information may be given and one would need to find a compromise or a trade-off among the requirements. It could also happen that a critical knowledge about the design is deliberately missing so that one would perform research and obtain the needed parameter, etc.

## 4 Classroom Design Approach

The traditional design method of above is good and is typically used for assignments and term projects. But how can one teach design in a classroom based setup? This is where we came up with an intuitive method that slowly builds several possible solutions to a design question using intuition as a primary source. In Table 1, we list five basic steps used toward this approach.

Let us walk briefly through an example, which literally could take up to two semesters to completely answer. Again, we must stress that our design focus is only aimed to an audience of communication engineering students. Therefore, we will only consider the corresponding part of interest as was explained and highlighted in Figure 2. Needless to say, that minor possible overlaps to adjacent fields such as DSP and antennas could still occur without loss of focus.

Table 1. Steps used in design based teaching

| Step #1 | Explain a design question in its simplest possible form ever. |
|---|---|
| Step #2 | Engage students for feedback and show open mindedness to all suggestions no matter how far-fetched it seems. |
| Step #3 | Build slowly the system and the theoretical knowledge required to comprehend it. |
| Step #4 | Add natural and manmade limitations to the system gradually one after the other [not all in one shot]. And explain ways to modify the system to cope with these changes. |
| Step #5 | Show parallel systems and compare them to identify advantages and disadvantages. |

So, we start the design question by explaining to students that physical layered telecommunications is all about sending binary [1 or 0] bits from some point A [a.k.a. transmitter] to some other geographical point B [a.k.a. receiver] as shown in the 1$^{st}$ image of Figure 4. Then, we ask the audience for possible suggestions in doing this. For example, a possible solution might be to store the data stream on a hard-drive and send it by courier to destination. But this approach is neither cost nor time effective. An alternate solution would be to use an electric wire as in the 2$^{nd}$ image. But the wire is not perfect, it has resistance, there will be loss or attenuation. Also, the receiver, which is made-up of electronic parts, adds thermal corruption, also known as noise, to the information bits. It turns out, that we could mathematically model this physical phenomenon by a so-called additive Gaussian noise as in the 3$^{rd}$ image. Then, we realize that the power of the noise signal is more dominant that that of the binary signal. In other words, the Signal to Noise Ratio [SNR] is a small value. So we figure out a way to change or map the binary information to some other signal such that the SNR is increased. This is referred to as modulation. We also explain that in communications there is always symmetry, so whatever we do on one side, the mirror reflection must also be applied on the other side, and hence a demodulator is used as in the 4$^{th}$ image. Even though a modulator/demodulator, or modem for short, decreases the effect of noise, it does not eliminate it completely. Therefore, some bits will still be detected in error, so a channel encoder is used to correct some of the errors encountered as in the 5$^{th}$ image. Now, we really have established the fundamentals of digital communications. It's interesting to note, that the design teaching up to this point takes roughly a semester to complete.

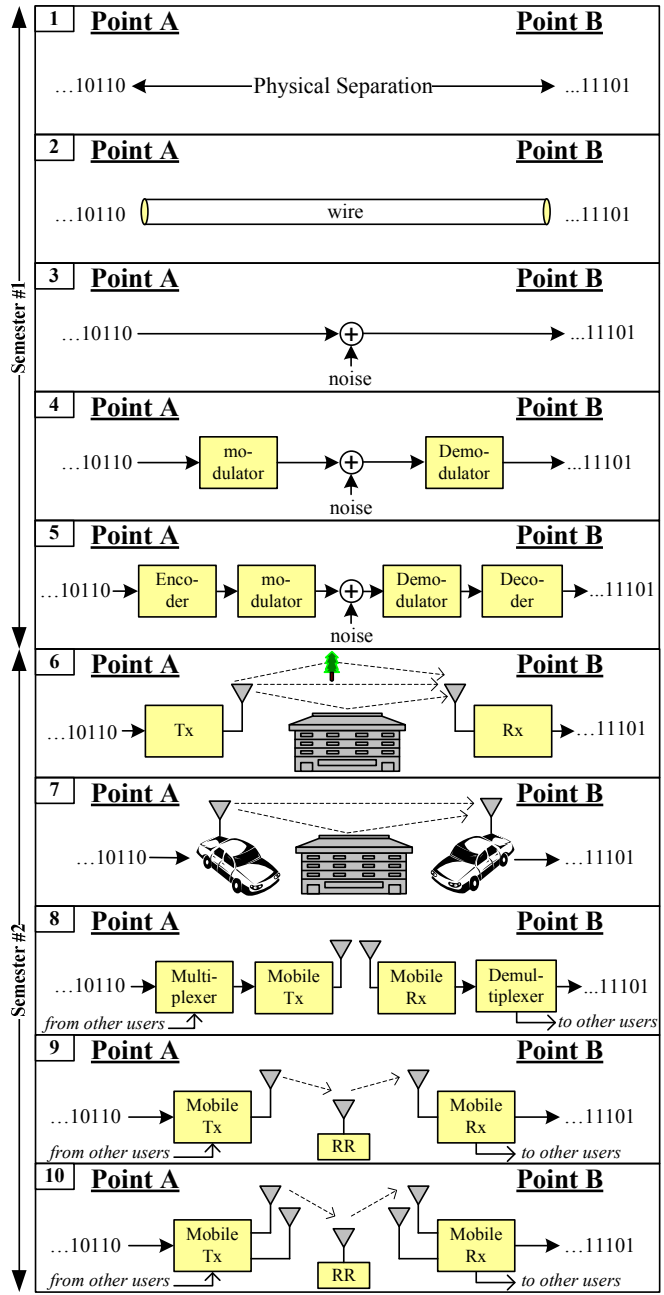

Figure 4. Design based teaching for wireless communications

In the following semester, we pickup where we left and start asking whether having a wire is always a good option. Is the use of a wireline practical? Cheap? How about the esthetics of a city? Say we have a lake, should we add protective isolation to the wire and put it into the lake? Or should we switch to optical communications and use a fibre optic? Or instead of this entire headache, maybe we could simply use a wireless antenna. Notice how here, we slowly build a socially responsible and convincing case for the gadget to be designed. Not based on mathematical equations, but really on intuition and common sense.

So now, we completely remove the wireline between transmitter and receiver and place an antenna on each side. But it just happens that we will run into other problems. The wave that travels from A to B will hit and gets reflected from various objects before it reaches destination. Some buildings are only made-up of concrete, others are entirely fabricated with glass, some buildings are small in size, others are skyscrapers, there are trees, pedestrians, cars that move, some days it rains, and in others it snows, etc. All the factors mentioned, among others, will have some effect and eventually need to be addressed one after the other as shown in the **6th** image. And we obviously do this to ensure a realistically good system model for proper design. Next, we ask students, what if we want to have the transmitter and receiver be mobiles and not fixed at all times. So, we start elaborating on the implications related to this as in the **7th** image. Then, we consider the communication of multiple users. Why should we reinvent the wheel? We could easily use the same system and merge the bits from other clients through multiplexing as in the **8th** image. Then, what if the distance between A and B is lengthy? This will for sure attenuate the power level of the signal, so maybe several Regenerative Repeaters [RR] could be used within the channel to refresh the bit stream as shown in the **9th** image. We could also extend this idea to long-distance intercontinental telecommunications, and show how a satellite could be used as an outer-space relay or RR. Then, we start adding several antennas to the system and notice how in this way the bits travel faster. In other words, one can download applications fast with the help of several antennas as oppose to only one, as shown in the **10th** image. And we could go on and on making the system better, more flexible, and complete.

Again, the point here isn't to explain to the reader how bits go from one place to another. But actually to show how gradually and systematically we build a complex design system by starting only with a simple design question. Now, it could happen that some instructors may argue and say that this approach is only possible in communications due to its modular nature. But, we truly think that any science or engineering field could be taught in class using this design based method, provided some thought and creativity is invested in it.

## 5 Analysis of Design

At each of the steps of say Figure 4, one needs to test the possible designs in order to comprehend and compare the various results. In the analysis of a system, standard methods are used. And, in general, two approaches are possible, either a low cost realization known as software simulation or a high cost hardware implementation. Figure 5, visually shows how the design based teaching of the previous section leads directly to practical simulation and implementation needed for testing, prototyping, and the evaluation of the system.

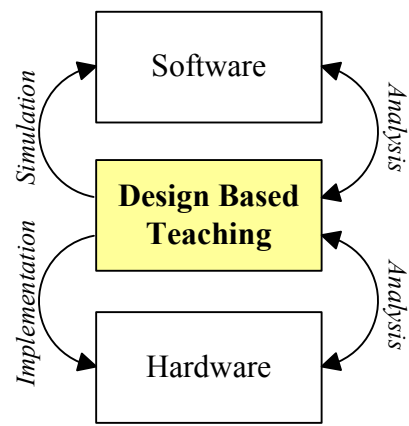

Figure 5. Analysis based on design teaching

## 7 Conclusion

In this paper, we started by explaining the importance of always connecting a narrow specialization to the big picture while demonstrating the interdependencies that exist between the various fields. Then, we showed the traditional method toward design and compared it to our classroom teaching approach. We also explained an example of how one is able to start with a basic design question and slowly mature into a sophisticated system using intuition as a principle source. Last, we briefly discussed how analysis also ties-in nicely to design based teaching.